\documentclass[nofootinbib,aps,11pt,amsmath,amssymb,a4paper,prd,onecolumn,showpacs]{revtex4-1}

\usepackage[english,american]{babel}
\usepackage[utf8]{inputenc}
\usepackage[T1]{fontenc}

\usepackage{graphicx}
\usepackage{color}

\usepackage{multirow}

\usepackage{units}

\allowdisplaybreaks

\usepackage{hyperref}
\hypersetup{
   pdfnewwindow=true,     
   colorlinks=true,       
   linkcolor=black,       
   citecolor=blue,        
   filecolor=black,       
   urlcolor=blue          
}

\begin{document}

\title{\Large {\bf{Simplified Dirac Dark Matter Models and Gamma-Ray Lines}}}
\author{Michael \surname{Duerr}}
\author{Pavel \surname{Fileviez P\'erez}}
\author{Juri \surname{Smirnov}}
\affiliation{\small{
Particle and Astroparticle Physics Division,\\
Max-Planck-Institut f\"ur Kernphysik,\\
Saupfercheckweg 1, 69117 Heidelberg, Germany}
}
\begin{abstract}
We investigate simplified dark matter models where the dark matter candidate is a Dirac fermion charged only under a new gauge symmetry. 
In this context one can understand dynamically the stability of the dark matter candidate and the annihilation through the new gauge boson is not velocity suppressed.
We present the simplest Dirac dark matter model charged under the local $B-L$ gauge symmetry. 
We discuss  in great detail the theoretical predictions for the annihilation into two photons, into the Standard Model Higgs and a photon, and into the $Z$ gauge boson 
and a photon. Our analytical results can be used for any Dirac dark matter model charged under an Abelian gauge symmetry.
The numerical results are shown in the $B-L$ dark matter model.  
We discuss the correlation between the constraints on the model from collider searches and dark matter experiments. 
\end{abstract}

\maketitle

\tableofcontents

\newpage

\section{Introduction}
\label{sec:introduction}
%
The existence of dark matter~(DM) in the Universe has motivated the particle physics community to investigate extensions of the Standard Model~(SM) of particle physics, with countless dark matter candidates on the market which need to be probed against experiments. There are basically three ways to search for these candidates:
one looks for possible signals in dark matter direct detection experiments, for gamma-ray lines and other signals from dark matter annihilation in indirect detection experiments, and for large missing energy signatures together with a mono-jet or a mono-photon at colliders. See Refs.~\cite{Jungman:1995df,Bergstrom:2000pn,Feng:2010gw,Bringmann:2012ez} for a detailed discussion of these possibilities. 

Most of the effort so far has been focused on studies of complete models and their dark matter candidates, as for example the neutralino in the Minimal Supersymmetric Standard Model. Also effective field theory (EFT) has been used to constrain the scale of the new physics, particularly using the data from the LHC run one. 
However, given the large center-of-mass energy of the LHC in its second  run, it is obvious that the EFT approach is prone to fail in large fractions of parameter space. Therefore, it is sensible to consider simplified dark matter models which can capture the main features of the dark matter sector with a limited number of parameters. For a discussion of simplified models see Refs.~\cite{Abdallah:2015ter,Buchmueller:2014yoa}.

A simplified model should contain a mediator and a dark matter candidate, and it should not violate generically any low energy observables.  
These criteria are necessary but not sufficient, since a naive model can result in misleading statements. We argue that the simplified model must be self consistent, i.e., not violate gauge invariance and be anomaly free. Only then is it possible to perform a full study and compute for example higher order processes leading to gamma-ray lines relevant for indirect dark matter searches. 

In this article we investigate in detail simplified models for Dirac dark matter, which have the following features:
\begin{itemize}
\item The dark matter is charged under a local gauge symmetry. This symmetry can be spontaneously broken at the low scale 
and a remnant discrete symmetry guarantees the dark matter stability. In the simplest case one has a local $U(1)^\prime$ symmetry broken to a $Z_2$ discrete symmetry. In the case of an unbroken gauge symmetry one can use the Stueckelberg mechanism and 
the dark matter stability is ensured by the choice of quantum numbers.
\item One can generate mass for the new gauge boson using the Stueckelberg or the Higgs mechanism. In the Stueckelberg scenario 
the dark matter candidate interacts only with the new neutral gauge boson in the theory. However, if one uses the Higgs mechanism the 
dark matter could also have interactions with the new physical Higgs boson.
\item The existence of a new gauge boson is key for the testability of the mechanism for dark matter stability. Therefore, in the ideal case the 
new gauge boson should define all the properties of the dark matter candidate. In particular, the main annihilation 
channels must proceed through the interaction between the dark matter and the new gauge boson. One can show that in this case
the dark matter annihilation through the new gauge boson is not velocity suppressed.
\end{itemize}

For a Dirac dark matter, we investigate in detail the relic density constraints, the predictions for direct detection and the dark matter annihilation channels 
producing monochromatic photons, $\bar{\chi} \chi  \to  Z^\prime  \to  \gamma \gamma,\ h \gamma,\ Z \gamma$.
We compute the one-loop generated vertices $Z^\prime  \gamma \gamma$, $Z^\prime h \gamma$ and $Z^\prime Z \gamma$ needed for the annihilation cross sections. We discuss all the technical details for the computation of the loop graphs and we stress the need to check
the Ward and Slavnov--Taylor identities to make sure the final results are correct. We point out that the effective coupling $Z^\prime  \gamma \gamma$ 
is possible only in models where the charged fermions inside the loop have an axial coupling to the $Z^\prime$. The effective couplings $Z^\prime h \gamma$ and $Z^\prime Z \gamma$ are always present 
when the Standard Model fermions are charged under the new gauge symmetry.  Our results can be used for the study of the gamma-ray lines in any theory with Dirac dark matter charged under a new gauge symmetry.

In order to illustrate the main results, we discuss the simplest possible self-consistent dark matter model which is most relevant for studying the connection between direct and indirect dark matter searches, as its annihilation cross section is not velocity suppressed.  
In this model, dark matter is charged under the $B-L$ gauge symmetry and one has only two annihilation channels into photons,
$\bar{\chi} \chi  \to  Z_{BL}  \to h \gamma, Z \gamma$. We discuss the parameter space for direct detection in agreement with the relic density and collider constraints, and we show the experimental limits on the indirect searches for the gamma-ray lines and $b\bar{b}$. We show that the interplay between the relic density, collider searches and indirect dark matter detection experiments sets non-trivial bounds on these simplified models. 
 
\section{Simplified Models}
We discuss models where the dark matter is a Dirac fermion $\chi$ charged only under a new gauge force. 
In this context we can understand why the dark matter is stable. For simplicity, we consider the case where one has an Abelian force, i.e., a $U(1)^\prime$. The part of the Lagrangian relevant for our discussion is 
\begin{equation}
 \mathcal{L} \supset - \frac{1}{4} F^{\prime \mu \nu} F_{\mu \nu}^\prime + i \bar{\chi}_L \gamma^\mu D_\mu \chi_L + i \bar{\chi}_R \gamma^\mu D_\mu \chi_R - \left( M_{\chi} \bar{\chi}_R {\chi}_L  + \text{h.c.} \right) 
 + \frac{1}{2} M_{Z^\prime}^2 Z^\prime_\mu Z^{\prime \mu},
\end{equation}   
where 
\begin{eqnarray}
F^\prime_{\mu \nu} & = & \partial_\mu Z_\nu^\prime - \partial_\nu Z_\mu^\prime, \\ 
D_\mu \chi_L &= & \left( \partial_\mu + i g^\prime n_L Z_\mu^\prime \right) \chi_L, \\
D_\mu \chi_R &=& \left( \partial_\mu + i g^\prime n_R Z_\mu^\prime \right) \chi_R.
\end{eqnarray}
Here we neglect the kinetic mixing between the new Abelian symmetry and $U(1)_Y$.
For the moment we do not discuss the anomaly cancellation and how the masses are generated but will address these issues later in a well-motivated model. Notice that in 
general the Standard Model fermions can be charged under the new symmetry such that new fermions are needed for anomaly cancellation. 

In these models the relevant interaction of the dark matter candidate $\chi = \chi_L + \chi_R$ to the $Z^\prime$ gauge boson is given by
\begin{equation}
 -i g^\prime \bar{\chi} \ \gamma^{\mu} \left(  n_L P_L + n_R P_R  \right)  \chi  Z_\mu^\prime,
\end{equation}
where we use the standard projection operators, $P_L = \frac{1}{2} (1-\gamma^5)$ and  $P_R = \frac{1}{2} (1+\gamma^5)$.
The interactions of all other fermions $f$ in the theory, the SM fermions or new fermions needed for anomaly cancellation, to the $Z^\prime$ can be parametrized as 
\begin{equation}
-i g^\prime \bar{f} \ \gamma^{\mu} \left(  g_V^f  -  g_A^f \gamma^5 \right)  f  Z_\mu^\prime.
\end{equation}
As usual, all charged fermions couple to the photon $A_\mu$ according to their electric charge $Q_f$,
\begin{equation}
 -i e Q_f \bar{f} \gamma^{\mu} f A_\mu,
\end{equation}
and the coupling of the fermions to the Standard Model $Z$ can be parametrized as 
\begin{equation}
 -i \frac{g_2}{\cos \theta_W} \bar{f} \gamma^\mu \left(g_L^f P_L + g_R^f P_R \right) f Z_\mu,
\end{equation}
where $g_2$ is the $SU(2)$ gauge coupling and $\theta_W$ is the Weinberg angle. 

\subsection{\texorpdfstring{$B-L$}{B-L} Dirac Dark Matter}
%
The local $B-L$ symmetry is anomaly free once we add three copies of right-handed neutrinos to the Standard Model particle content. It is well known that 
this symmetry could play a major role in neutrino physics. Here we focus on a very simple model with Dirac 
dark matter charged under $B-L$. The relevant part of the Lagrangian is given by
\begin{equation}
 \mathcal{L}_{BL} \supset  i \bar{\chi} \gamma^\mu D_\mu \chi -  M_{\chi} \bar{\chi} \chi + \frac{1}{2} M_{Z_{BL}}^2 Z_{BL \mu} Z^{\mu}_{BL}, 
\end{equation}  
where $D^\mu \chi = \left( \partial^\mu + i g_{BL} \ n \ Z_{BL}^{ \mu} \right) \chi$, $n \neq \pm 1$, and $\chi=\chi_L + \chi_R$. One can generate the gauge boson mass through the Higgs mechanism or the Stueckelberg mechanism. Let us discuss both cases here:

\begin{itemize}

\item \textit{Stueckelberg Mechanism:}
The mass of the $B-L$ gauge boson can be generated through the Stueckelberg mechanism as discussed in Ref.~\cite{Nath}:
\begin{equation}
 {{\cal L}}_{BL} \supset    \frac{1}{2} \left(  M_{Z_{BL}} Z_{BL \mu} +  \partial_\mu \sigma \right) \left( M_{Z_{BL}} Z_{BL}^{\mu} +  \partial^\mu \sigma \right) - \left(  Y_\nu \ \bar{\ell}_L \tilde{H} \nu_R + \textrm{h.c.} \right),
\end{equation}  
where the gauge transformations are given by
\begin{equation}
\delta Z_{BL}^{\mu} = \partial^\mu \lambda \text{ and } \delta \sigma = - M_{Z_{BL}}  \lambda.
\end{equation}
In this case the neutrinos are Dirac fermions because the $B-L$ symmetry is never broken, and the dark matter stability is a result of the choice of the quantum number for the dark matter candidate.
\item \textit{Higgs Mechanism:}
One can generate the mass for the $B-L$ gauge boson through the Higgs mechanism and at the same time we can generate masses for the SM neutrinos through the see-saw mechanism~\cite{TypeI-1,TypeI-2,TypeI-3,TypeI-4,TypeI-5} via the following interactions:
\begin{equation}
-  {{\cal L}}_{\nu} = Y_\nu \ \bar{\ell}_L \tilde{H} \nu_R + \lambda_R \nu_R \nu_R S_{BL} + {\rm{h.c.}}
\end{equation}
Here $S_{BL}$ is a Standard Model singlet and has  $B-L$ charge two. Notice that if the $B-L$ charge of the new Higgs is different from two, the neutrinos will be Dirac fermions. After the $U(1)_{B-L}$ is broken, there is a remnant $Z_2$ symmetry which is the reason for the dark matter stability.
\end{itemize}

These simple models have only four relevant parameters for the dark matter study: the gauge coupling $g_{BL}$, the dark matter mass $M_\chi$, the gauge boson mass 
$M_{Z_{BL}}$, and $n$ the $B-L$ charge of the dark matter candidate. The relevant interactions needed to compute 
the dark matter annihilation channels are
\begin{eqnarray}
& & -i g_{BL} n \ \bar{\chi}  \gamma_{\mu}  \chi  Z_{BL}^{\mu} \qquad  {\text{and}} \qquad 
 -i g_{BL} n_{BL}^f \ \bar{f}  \gamma_{\mu}  f  Z_{BL}^{\mu},
\end{eqnarray}
where $n_{BL}^f$ is the $B-L$ charge of the Standard Model fermion $f$.
%
\subsubsection{Relic Density}\label{sec:relicDensity}
%
The $B-L$ dark matter candidate $\chi$ can annihilate into all the Standard Model particles and the $B-L$ gauge boson $Z_{BL}$. 
Therefore, one can have the annihilation channels
\begin{displaymath}
\bar{\chi}  \chi \ \to \ \bar{q} q, \ \bar{\ell} \ell, \ \bar{\nu} \nu, \ Z_{BL} Z_{BL}
\end{displaymath}
in both the Stueckelberg and the Higgs scenario discussed above. In the Higgs scenario, one has the additional annihilation to right-handed neutrinos. There are two main regimes for our study:
\begin{itemize}
\item $M_\chi < M_{Z_{BL}}$: when the dark matter candidate is lighter than the $B-L$ gauge boson, we have the following channels,
\begin{displaymath}
\bar{\chi}  \chi \ \to \ Z_{BL}^\ast \  \to \ \bar{q} q, \bar{\ell} \ell, \bar{\nu} \nu, (\nu_R \nu_R). 
\end{displaymath} 
\item $M_{Z_{BL}} < M_{\chi}$: when the dark matter candidate is heavier than the $B-L$ gauge boson,
one has a new open channel which is not velocity suppressed,
\begin{displaymath}
\bar{\chi}  \chi \ \to \ Z_{BL} Z_{BL}.
\end{displaymath} 
\end{itemize}
In order to test this model at the collider, the invisible decay $Z_{BL} \to \bar{\chi} \chi$ is crucial to establish the connection between the existence of the new gauge boson and the dark matter candidate.
Therefore, we focus on the first regime. 

\begin{figure}[t]
 \includegraphics[width=0.48\linewidth]{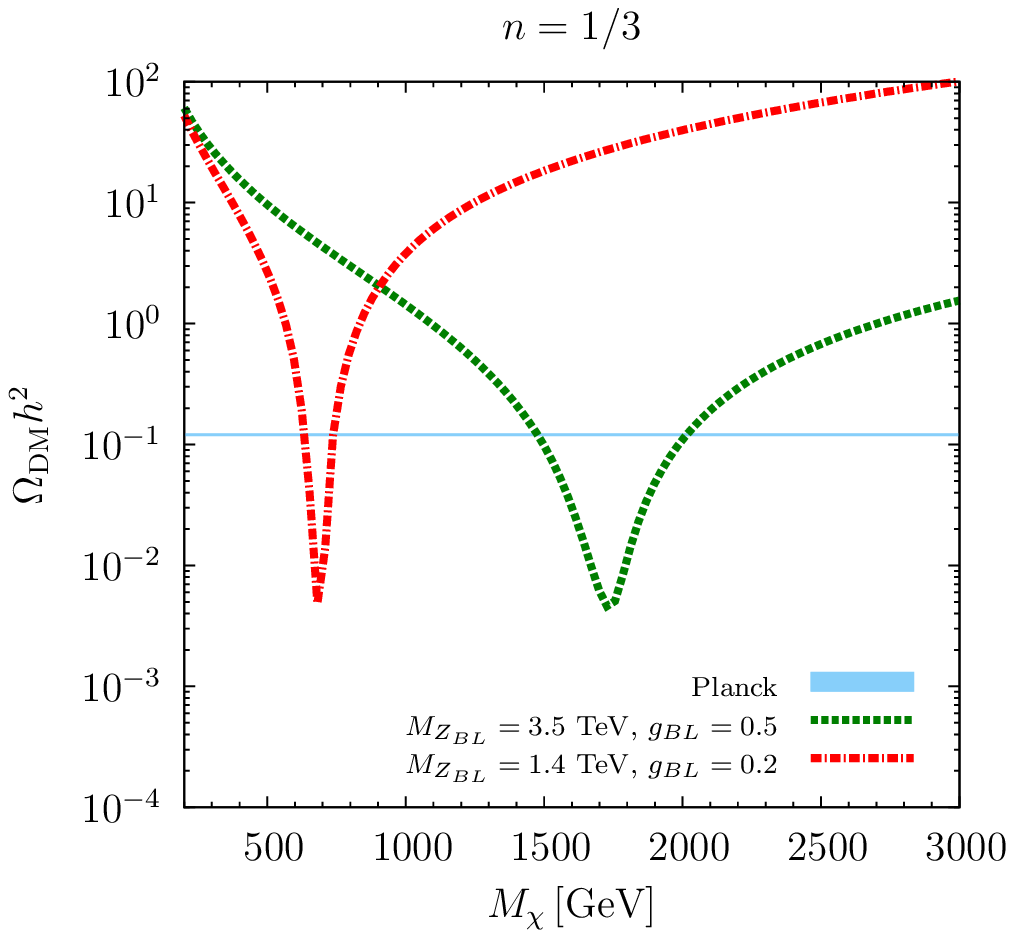}
 \includegraphics[width=0.48\linewidth]{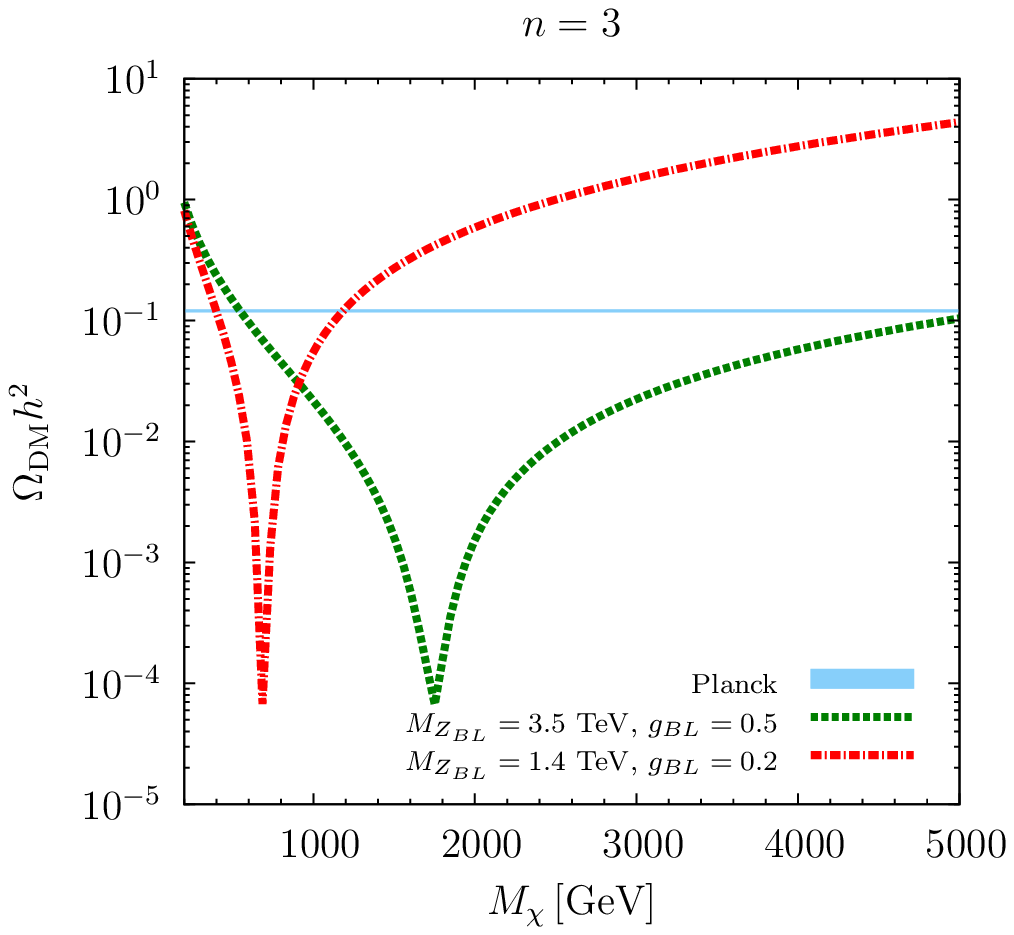}
 \caption{Dark matter relic density $\Omega_\text{DM} h^2$ vs.\ the dark matter mass $M_\chi$ for different choices of the dark matter $B-L$ quantum number $n$. In the left panel we use $n = 1/3$, while in the right panel we use $n = 3$, and we give the relic density for two choices of the mass of the $Z_{BL}$ and the gauge coupling $g_{BL}$ that fulfill $M_{Z_{BL}}/g_{BL} = \unit[7]{TeV}$. The thin blue band corresponds to the currently allowed relic dark matter density measured by the Planck collaboration, $\Omega_\text{DM} h^2 = 0.1199 \pm 0.0027$~\cite{Ade:2013zuv}. \label{fig:relicDensity} } 
\end{figure}

The annihilation cross section for $\bar{\chi} \chi \ \to \ Z_{BL}^\ast \  \to \ \bar{f} f$ is given by
\begin{equation}
\label{eq:AnnihilationCrossSection}
\sigma (\bar{\chi} \chi \to  Z_{BL}^\ast   \to  \bar{f} f) = \frac{N_c^{f} (n_{BL}^{f})^2 g_{BL}^4 n^2 }{12 \pi s} \frac{\sqrt{s - 4 M_{f}^2}}{\sqrt{s - 4 M_\chi^2}} \frac{\left( s + 2 M_\chi^2\right) \left( s + 2 M_{f}^2 \right)}{ \left[ (s- M_{Z_{BL}}^2)^2 + M_{Z_{BL}}^2 \Gamma_{Z_{BL}}^2 \right]}.
\end{equation}
Here $N_c^f$ is the color factor of the fermion $f$ with mass $M_f$, $s$ is the square of the center-of-mass energy, and $\Gamma_{Z_{BL}}$ is the total decay width of the $Z_{BL}$ gauge boson.
In order to compute the relic density we use the analytic approximation~\cite{Gondolo:1990dk}
\begin{equation}
\label{eq:relicdensity}
\Omega_\text{DM} h^2 = \frac{\unit[2.14 \times 10^{9}]{GeV}^{-1}}{J(x_f) \sqrt{g_\ast} \ M_\text{Pl}},
\end{equation}
where $M_\text{Pl}=\unit[1.22 \times 10^{19}]{GeV}$ is the Planck scale, 
$g_\ast$ is the total number of effective relativistic degrees of freedom at the time of freeze-out, 
and the function $J(x_f)$ reads as
\begin{equation}
J(x_f)=\int_{x_f}^{\infty} \frac{ \langle \sigma v \rangle (x)}{x^2} dx.
\end{equation}
The thermally averaged annihilation cross section times velocity $\langle \sigma v \rangle$ is a function of $x=M_\chi/T$, and is given by
\begin{equation}
 \langle\sigma v\rangle (x) = \frac{x}{8 M_\chi^5 K_2^2(x)} \int_{4 M_\chi^2}^\infty \sigma \times ( s - 4 M_\chi^2) \ \sqrt{s} \ K_1 \left(\frac{x \sqrt{s}}{M_\chi}\right) ds,
\end{equation}
where $K_1(x)$ and $K_2(x)$ are the modified Bessel functions.
The freeze-out parameter $x_f$ can be computed using
\begin{equation}
x_f= \ln \left( \frac{0.038 \ g \ M_\text{Pl} \ M_\chi \ \langle\sigma v\rangle (x_f) }{\sqrt{g_\ast x_f}} \right),
\end{equation}
where $g$ is the number of degrees of freedom of the dark matter particle.  In Fig.~\ref{fig:relicDensity} we show the numerical predictions for the relic density vs.\ the dark matter mass for two values of $n$.
In the left (right) panel we show the results for $n=1/3 \, (3)$ when $M_{Z_{BL}}/g_{BL} = \unit[7]{TeV}$, which is in agreement with the collider bounds~\cite{Carena}. As expected, for small values of the gauge coupling one needs 
to rely on the resonance to achieve the right relic density. However, generically one can be far from the resonance and in agreement with relic density constraints.

\subsubsection{Direct Detection}
The direct detection constraints must be considered in order to understand which are the allowed values of the input parameters in this theory. 
The elastic spin-independent nucleon--dark matter cross section is given by
\begin{equation}
\sigma_{\chi N}^\text{SI} = \frac{M_N^2 M_\chi^2}{ \pi (M_N + M_\chi)^2} \frac{g_{BL}^4}{M_{Z_{BL}}^4} n^2,
\end{equation}
where $M_N$ is the nucleon mass. Notice that $\sigma_{\chi N}^\text{SI}$ is independent of the matrix elements. The cross section can be rewritten as 
\begin{equation}
 \sigma_{\chi N}^\text{SI} (\text{cm}^2) = 12.4 \times 10^{-41} \left( \frac{\mu}{\unit[1]{GeV}}\right)^2 \left( \frac{\unit[1]{TeV}}{r_{BL}}\right)^4 n^2 \ \text{cm}^2,
\end{equation}
where $\mu = M_N M_\chi / (M_N + M_\chi)$ is the reduced mass and $r_{BL} = M_{Z_{BL}}/g_{BL}$.

In our case $M_\chi \gg M_N$, and using the collider lower bound $M_{Z_{BL}}/ g_{BL} > \unit[6]{TeV}$~\cite{Carena} one finds an upper bound on the elastic spin-independent nucleon--dark matter cross section given by
\begin{equation}
\sigma_{\chi N}^\text{SI} < 9.57 \times 10^{-44}  n^2 \ \text{cm}^2,
\end{equation}
for a given value of $n$. There is also a simple way to find a lower bound on the spin-independent cross section. The minimal value of the gauge coupling 
$g_{BL}$ in agreement with relic density constraints corresponds to the case when one sits on the resonance, i.e., $M_{Z_{BL}}=2 M_\chi$. Therefore, the lower 
bound on the cross section for a given value of the dark matter mass is given by
\begin{equation}
 \sigma_{\chi N}^\text{SI} > 7.75 \times 10^{-42}  \left( \frac{\unit[1]{TeV}}{M_\chi}\right)^4 (g_{BL}^{\text{min}})^4 n^2 \ \text{cm}^2.
\end{equation}

\begin{figure}[t]
 \includegraphics[width=0.48\linewidth]{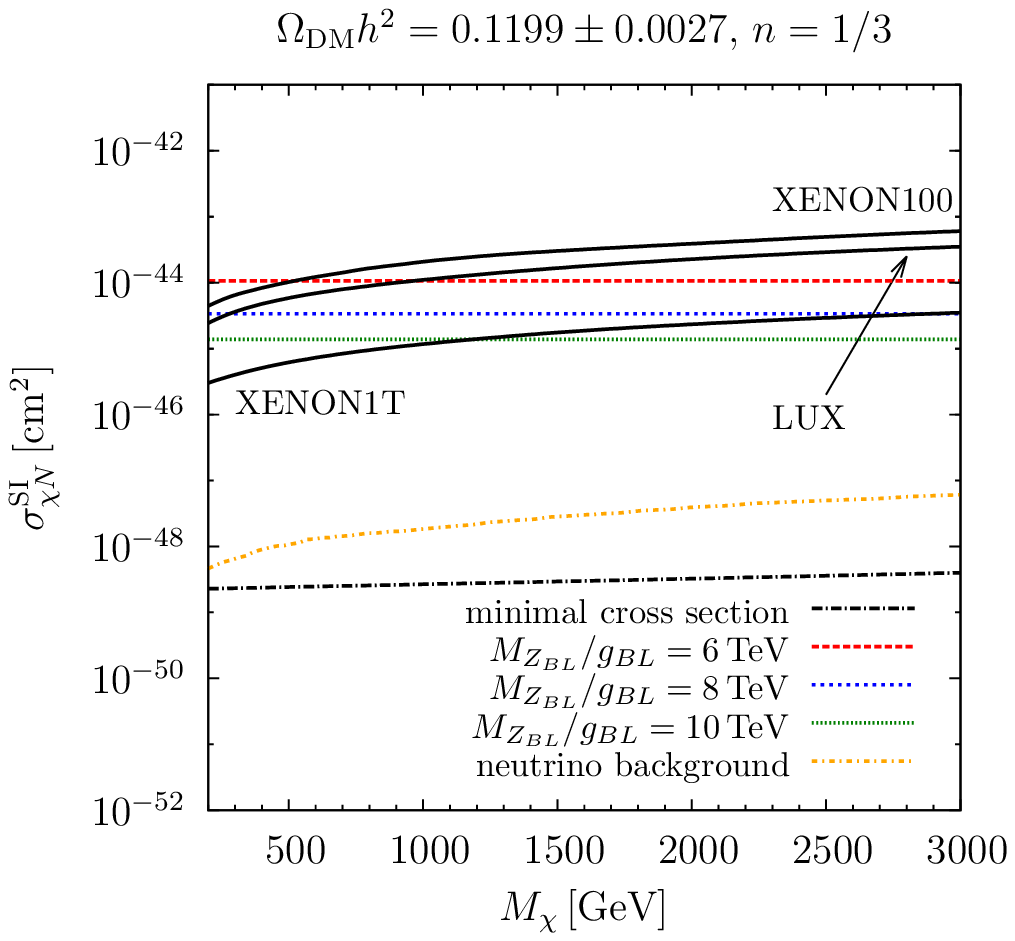}
 \includegraphics[width=0.48\linewidth]{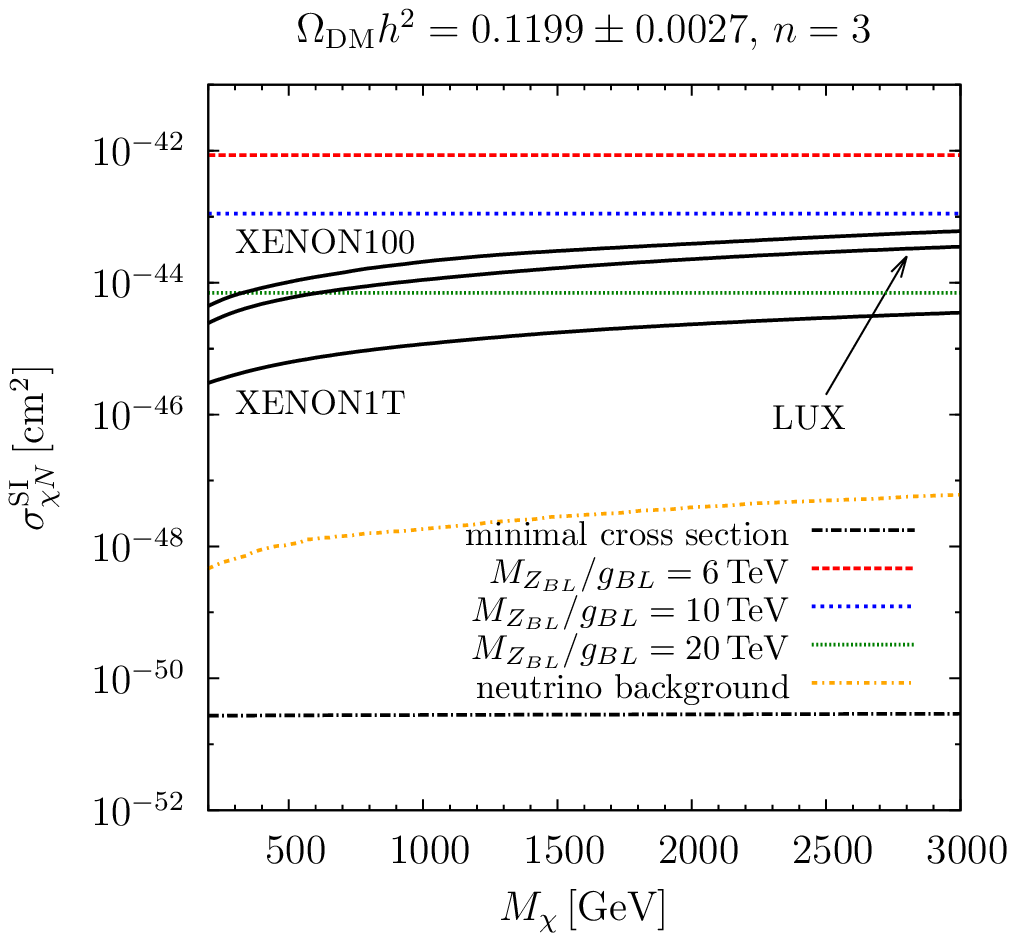}
 \caption{Direct detection cross section $\sigma_{\chi N}^\text{SI}$ vs.\ the dark matter mass $M_\chi$ compatible with the relic density constraints for $n=1/3$ (left panel) and $n=3$ (right panel). The colored dashed lines show $\sigma_{\chi N}^\text{SI}$ for different choices of $M_{Z_{BL}}/g_{BL}$ compatible with current collider limits. The black dash-dotted line shows the minimal direct detection cross section. We show the LUX~\cite{Akerib:2013tjd} and XENON100~\cite{Aprile:2012nq} constraints, as well as the prospects for XENON1T~\cite{Aprile:2012zx}. The orange dash-dotted line shows the coherent neutrino scattering background~\cite{Billard:2013qya}. \label{fig:directDetection} } 
\end{figure}

In Fig.~\ref{fig:directDetection} we show the numerical predictions for the direct detection cross section $\sigma_{\chi N}^\text{SI}$ vs.\ the dark matter mass $M_\chi$ compatible with the relic density constraints. 
The colored dashed lines show the values of $\sigma_{\chi N}^\text{SI}$ for different choices of $M_{Z_{BL}}/g_{BL}$ compatible with current collider limits. 
The black dash-dotted line shows the minimal direct detection cross section. We show the bounds from the LUX~\cite{Akerib:2013tjd} and XENON100 experiments~\cite{Aprile:2012nq}, as well as the prospects for XENON1T~\cite{Aprile:2012zx}. One can see that for $n=1/3$, the scenario for $M_{Z_{BL}}/g_{BL} = \unit[6]{TeV}$ is allowed by the LUX experiment for a large part of the parameter space. However, for the case $n=3$, the ratio $M_{Z_{BL}}/g_{BL}$ needs to be larger than \unit[20]{TeV} in order to satisfy the experimental bounds. 
Therefore, only the scenario when $n=1/3$ could be tested at the LHC.  
Unfortunately, the minimal value of the cross section is below the neutrino background~\cite{Billard:2013qya} and it is very difficult to test this part of the parameter space in the current direct detection experiments.

\subsection{Upper Bound on the Dark Matter Mass}
In this section we show that in these simple models it is possible to derive an upper bound on the dark matter mass. 
We focus on the case when the $Z^\prime$ is heavier than the dark matter because only then one can test the main properties of these models.
The argument is based on the observational requirement that the relic density of dark matter produced in the freeze-out must not overclose the 
Universe. Today, we know that $\Omega_\text{DM} h^2 \leq 0.12$ and since the relic density scales 
as  $\Omega_\text{DM} h^2 \propto 1/ {\langle  \sigma v \rangle}$, one has a lower bound 
on the annihilation cross section
\begin{align}
\langle  \sigma v \rangle \gtrsim \langle  \sigma v \rangle_0 \approx 3 \times 10^{-26} \text{cm}^3/\text{s}\,.
\end{align}
Here $\langle \sigma v \rangle_0 $ is the minimal value of the cross section compatible with observations. 
In order to guarantee the validity of a given theory we have to make sure that the maximal value of the cross section in the theory obeys the condition  $\langle \sigma v \rangle_\text{max} \geq \langle \sigma v \rangle_0$. This is a necessary condition, since if it is not fulfilled there is no parameter choice in the model which can make it compatible with observations. 

The corresponding cross section has the following structure,
\begin{align}
\sigma v (g^\prime,  M_{Z^\prime}, M_\chi)= \frac{c_1 (g^\prime)^4 M_\chi^2}{(4 M_\chi^2-M_{Z^\prime}^2 )^2 + M_{Z^\prime}^2 \Gamma_{Z^\prime}^2 } \text{  with  } \Gamma_{Z^\prime} = c_2 (g^\prime)^2  M_{Z^\prime}\,.
\end{align}
It is obvious that the maximum of this expression given a fixed value of $g^\prime$ is realized when $2 M_\chi =  M_{Z^\prime}$. 
Examining the functional dependence of the obtained expression one finds that 
\begin{align}
\langle \sigma v \rangle_\text{max} = \frac{c_1}{16 c_2^2 M_\chi^2} \gtrsim  3 \times 10^{-26} \text{cm}^3/\text{s}\,.
\end{align}

In the model with local $B-L$ and $n = 1/3$, this leads to an upper bound on the mass of 
\begin{equation}
M_\chi \lesssim \unit[12.3]{TeV} \quad \text{and} \quad M_{Z_{BL}} \lesssim \unit[24.6]{TeV}.
\end{equation} 
Note that this bound is conservative because at the resonance one would need to perform the full average as discussed in Sec.~\ref{sec:relicDensity}.
This bound is useful to understand the possibility to test this type of model.

\section{Gamma-Ray Lines}
%
In this section, we discuss the predictions for gamma-ray lines in detail. 
First, we give the general results for any simplified model with an Abelian gauge symmetry and a corresponding $Z^\prime$, then we move on to study numerically the predictions for the minimal $B-L$ model discussed before. 
\subsection{Loop-Induced Couplings}

\begin{figure}[t]
 \centering
 \includegraphics[width=0.4\linewidth]{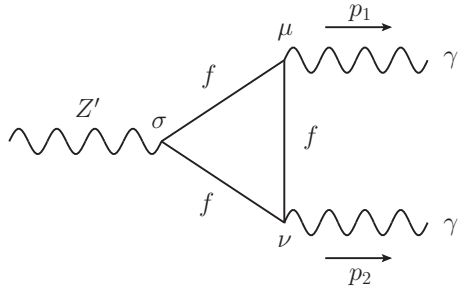}
 \caption{Coupling $Z^\prime \gamma \gamma$ generated by a loop of fermions $f$ charged under $U(1)^\prime$ 
 and carrying electric charge. In the calculation, the crossed diagram has to be taken into account.\label{fig:Zprimegammagamma}}
\end{figure}

In order to understand the predictions for the dark matter annihilation into photons, we need to compute the loop-induced effective interactions $Z^\prime \gamma \gamma$, $Z^\prime h \gamma$, and $Z^\prime Z \gamma$ shown in Figs.~\ref{fig:Zprimegammagamma}--\ref{fig:ZprimeZgamma}.

\begin{itemize}
\item {\textit{$Z^\prime \gamma \gamma$ coupling}}:
The coupling $\delta \Gamma_{Z^\prime \gamma \gamma}^{\mu \nu \sigma}$ is generated by a loop of electrically charged fermions $f$ charged also under $U(1)^\prime$, see Fig.~\ref{fig:Zprimegammagamma}, and is given by
\begin{align}\label{eq:Zprimegammagamma}
\delta \Gamma_{Z^\prime \gamma \gamma}^{\mu \nu \sigma}&= \sum_f \left[  \epsilon^{\mu \nu \sigma \alpha} (p_1-p_2)_\alpha - \frac{2}{s} \epsilon^{\mu \sigma \alpha \beta} \ p_{1\alpha} p_{2\beta} p_{1}^{\nu} 
+ \frac{2}{s}  \epsilon^{\nu \sigma \alpha \beta} \ p_{1\alpha} p_{2\beta} p_{2}^\mu \right] A_1^f \nonumber \\
&\quad + \sum_f \left( \epsilon^{\mu \sigma \alpha \beta} p_{1 \alpha} p_{2 \beta} p_2^\nu - \epsilon^{\nu \sigma \alpha \beta} \ p_{1\alpha} p_{2\beta} p_{1}^\mu \right) A_4^f 
 + \sum_f \epsilon^{\mu \nu \alpha \beta} p_{1 \alpha} p_{2 \beta} (p_1+p_2)^\sigma A_7^f.
\end{align}
Since we are interested in processes with two external photons, one has
 $p_1^\mu \epsilon^\ast_\mu(p_1) = p_2^\nu \epsilon^\ast_\nu (p_2) = 0$
and the terms in the second line of Eq.~\eqref{eq:Zprimegammagamma} proportional to $A_4^f$ do not contribute to the amplitude. The relevant coefficient functions are given by
\begin{align}
  A_1^f &=  \frac{e^2 Q_f^2 g' g_A^f N_c^f}{4 \pi^2} \left[ 3 + \Lambda(s, M_f, M_f) + 2 M_f^2 \ C_0(0, 0, s; M_f, M_f, M_f)\right],\\
  A_7^f &= \frac{e^2 Q_f^2 g' g_A^f N_c^f}{2 \pi^2 s} \left[ 2 + \Lambda(s, M_f, M_f) \right].
 \end{align}
Notice that this coupling can be generated only if the axial coupling $g_A^f$ of the fermions in the loop to the $Z^\prime$ is different from zero. See Appendix~\ref{app:loopFunctions} for the explicit form of the loop functions $\Lambda(s, M_f, M_f)$ and $C_0(0, 0, s; M_f, M_f, M_f)$. In the above equations $s=(p_1+p_2)^2$, $M_f$ is the fermion mass, and $N_c^f$ is the color factor and $Q_f$ the electric charge of the fermion. We checked that the Ward identities 
$\delta \Gamma_{Z^\prime \gamma \gamma}^{\mu \nu \sigma} \ p_{1 \mu}=\delta \Gamma_{Z^\prime \gamma \gamma}^{\mu \nu \sigma} \ p_{2 \nu}=0$
are satisfied.

\begin{figure}[t]
 \centering
 \includegraphics[width=0.4\linewidth]{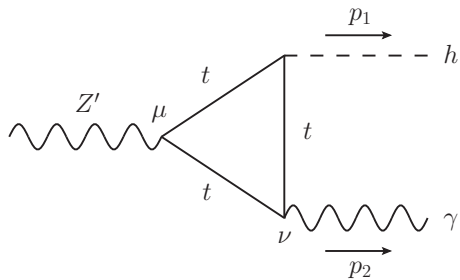}
 \caption{Coupling $Z^\prime h \gamma$ generated by a top loop. 
 In the calculation, the crossed diagram has to be taken into account.}
 \label{fig:Zprimehgamma}
\end{figure}

\item {\textit{$Z^\prime h \gamma$ coupling}}:
In Fig.~\ref{fig:Zprimehgamma} we show the coupling $Z^\prime h \gamma$ generated by a top loop. The other Standard Model fermions have smaller Yukawa couplings and their contributions are therefore negligible. For the top quark $g_A^t=0$, and the coupling is given by
\begin{equation}
 \delta \Gamma_{Z^\prime h \gamma}^{\mu \nu} = C_{Z^\prime h \gamma}  \left[ 2 p_1^\nu p_2^\mu + \left(M_h^2-s \right) g^{\mu\nu}\right] + p_2^\nu \tilde{C}_{Z^\prime h \gamma}^\mu,
\end{equation}
where
\begin{multline}\label{eq:hgammaCoefficient}
 C_{Z^\prime h \gamma} =  (-i) \frac{3}{4 \pi^2}  \frac{g^\prime g_V^t e Q_t M_t^2}{v_0 (s-M_h^2)^2} 
 \left\{ 2 s \left[ \Lambda(M_h^2,M_t,M_t) - \Lambda(s,M_t,M_t) \right] \right.   \\
+ \left. (M_h^2 -s ) \left[ 2 + (s + 4 M_t^2 - M_h^2 ) C_0(0,M_h^2,s;M_t,M_t,M_t) \right] \right\}.
\end{multline}
Here $v_0$ is the vacuum expectation value of the Standard Model Higgs. Notice that for processes with an external photon, the $\tilde{C}_{Z^\prime h \gamma}$ term does not contribute to the amplitude. 
See Appendix~\ref{app:loopFunctions} for the explicit form of the loop functions. 
For this vertex, one can show that the Ward identity is satisfied, $\delta \Gamma_{Z^\prime h \gamma}^{\mu \nu} \ p_{2 \nu}=0$.

\begin{figure}[t]
 \centering
 \includegraphics[width=0.4\linewidth]{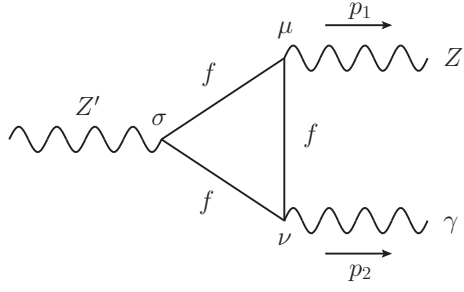}
 \caption{Coupling $Z^\prime Z \gamma$ generated by a loop of fermions charged under $U(1)^\prime$ 
 and carrying electric charge. In the calculation, the crossed diagram has to be taken into account.}
 \label{fig:ZprimeZgamma}
\end{figure}

\item\textit{$Z^\prime Z \gamma$ coupling:}
The coupling between the $Z^\prime$, the photon, and the $Z$ can be generated at one-loop level as shown in Fig.~\ref{fig:ZprimeZgamma}. This coupling can be written as
\begin{equation}
\delta \Gamma^{\mu \nu \sigma}_{Z^\prime Z \gamma}=- \frac{g^\prime g_2 e Q_f N_c^f}{16 \pi^2 \cos \theta_W} B^{\mu \nu \sigma},
\end{equation}
where
\begin{align}\label{eq:ZprimeZgamma}
B^{\mu \nu \sigma} &= \sum_f \bigg\{ \epsilon^{\mu \nu \sigma \alpha} p_{2 \alpha} B_2^f + B_3^f \left[ \frac{1}{2} (M_Z^2 - s) \epsilon^{\mu \nu \sigma \alpha} p_{1\alpha} + \epsilon^{\mu \sigma \alpha \beta} p_{1 \alpha} p_{2 \beta} p_{1}^{ \nu} \right]  \nonumber \\
&\qquad \quad +   \epsilon^{\mu \sigma \alpha \beta} p_{1 \alpha} p_{2 \beta} p_2^\nu B_4^f + \epsilon^{\nu \sigma \alpha \beta} p_{1\alpha} p_{2\beta} p_{1}^\mu B_5^f \\ 
 &\qquad \quad +  \epsilon^{\nu \sigma \alpha \beta} p_{1 \alpha} p_{2 \beta} p_2^\mu B_6^f + \epsilon^{\mu \nu \alpha \beta} p_{1 \alpha} p_{2 \beta} \left( p_{1}^\sigma B_7^f + p_{2}^\sigma B_8^f \right)\bigg\} . \nonumber
\end{align}
For processes with external $Z$ and $\gamma$ only, the terms in the second line of Eq.~\eqref{eq:ZprimeZgamma} proportional to $B_4^f$ and $B_5^f$ do not contribute and the relevant $B_i^f$ functions are given by
\begin{align}
 B_2^f &=  \frac{4}{\left( M_Z^2 -s  \right)} 
  \left\{ \left( A^f g_A^f + B^f g_V^f \right) \right. \nonumber \\ 
  &\quad \times \left[ 3 \left( M_Z^2 - s + M_{Z}^2 \Lambda(M_Z^2,M_f, M_f) \right) - ( 2 M_{Z}^2 + s) \Lambda(s,M_f, M_f) \right]   \\
  & \quad +  \left. \left[ B^f g_V^f (M_Z^2 - 2 M_f^2) + A^f g_A^f (M_Z^2 + 2 M_f^2) \right] (M_Z^2- s) C_{0} (0, M_Z^2, s; M_f, M_f, M_f)  \right\}, \nonumber\\
 B_3^f &=  - \frac{8}{\left( s- M_Z^2  \right)^2} 
  \left( A^f g_A^f + B^f g_V^f \right)  \left[ s \Lambda(M_Z^2,M_f, M_f) - s\Lambda(s,M_f, M_f) \right. \nonumber \\
  & \quad + \left.\left( M_Z^2 - s \right)\ \left( 1 + 2 M_f^2 C_0(0, M_Z^2, s; M_f, M_f, M_f) \right) \right], \\
 B_6^f &= \frac{8}{\left( s - M_Z^2 \right)^2} 
  \left( A^f g_A^f + B^f g_V^f \right)  \left\{ 3 M_Z^2 \Lambda(M_Z^2, M_f, M_f) - (2 M_Z^2 + s) \Lambda(s, M_f, M_f) \right. \nonumber \\ 
  & \quad + \left. (M_Z^2 - s) \left[3 + (2 M_f^2 + M_Z^2) C_0(0, M_Z^2, s; M_f, M_f, M_f) \right] \right\},\\
 B_7^f &= \frac{8}{\left( s - M_Z^2 \right)} 
  \left( A^f g_A^f + B^f g_V^f \right)  \left[ \Lambda(M_Z^2,M_f, M_f) - \Lambda(s,M_f, M_f)\right],\\
 B_8^f &= \frac{8}{\left( s - M_Z^2  \right)^2} 
  \left( A^f g_A^f + B^f g_V^f \right) \left[2 M_Z^2 \Lambda(M_Z^2, M_f, M_f) - (M_Z^2 + s) \Lambda(s, M_f, M_f) \right. \nonumber\\
 & \quad   \left. + (M_Z^2 - s) \left(2 + M_Z^2 C_0(0, M_Z^2, s; M_f, M_f, M_f)\right)\right]. 
\end{align}
Here the coefficients $A^f$ and $B^f$ are given by
\begin{eqnarray}
A^f&=& \frac{1}{2} \left( g_L^f + g_R^f\right) \text{ and }
B^f = \frac{1}{2} \left( g_L^f - g_R^f\right).
\end{eqnarray}
Using the Ward identity one can write $B_1^f$ as a function of $B_3^f$: 
\begin{equation}
B_1^f=-\frac{1}{2} \left( s - M_Z^2 \right) B_3^f.
\end{equation}
In the case of a spontaneously broken gauge symmetry, one has to use the Slavnov--Taylor identity~\cite{Slavnov:1972fg,Taylor:1971ff}
\begin{equation}
-i p_{1\mu} \delta \Gamma^{\mu \nu \sigma}_{Z^\prime Z \gamma}= M_Z \delta \Gamma^{ \nu \sigma}_{Z^\prime A \gamma},
\end{equation}
which is the generalization of the Ward identity. Here, $A$ is the Goldstone boson. From this identity, one finds the relation
\begin{equation}
B_2^f = - M_Z^2 B_5^f - \frac{1}{2} (s - M_Z^2) B_6^f - 16 B^f g_V^f M_f^2 C_0 (0, M_Z^2, s; M_f, M_f,M_f). 
\end{equation}
We have checked that these identities are satisfied in our calculations. Since for our case we find $B_5^f=0$, one can write
\begin{align}
 B_2^f &= - \frac{1}{2} (s - M_Z^2) B_6^f - 16 B^f g_V^f M_f^2 C_0 (0, M_Z^2, s; M_f, M_f,M_f) \nonumber \\
       &\equiv  - \frac{1}{2} (s - M_Z^2) B_6^f - C_0^f.
\end{align}

It can easily be checked that the following additional relation between the coefficients holds,
\begin{equation}
 B_7^f = B_8^f - B_6^f - B_3^f. 
\end{equation}
These relations are very useful to cross-check the results and simplify the final expressions for the cross sections.
We have used Package-X~\cite{Hiren} to perform all one-loop calculations and have cross-checked the results.
\end{itemize}

Using the above calculations for the loop-induced couplings, we compute the dark matter annihilation cross sections for the different channels:
\begin{itemize}

\item ${\bar{\chi} \chi \to \gamma \gamma}$: 
The amplitude for the dark matter annihilation into two photons is given by
\begin{multline}
\label{eq:ZBgammagammamatrixelement}
\left| \overline{\mathcal{M}} ({\bar{\chi} \chi \to \gamma \gamma}) \right|^2 = \frac{\left|\sum_f \left( A_7^f - \frac{2}{s} A_1^f \right) \right|^2 \left(n_R - n_L \right)^2 (g^\prime)^2 M_\chi^2 s^3 }{4 M_{Z^\prime}^2 \left( M_{Z^\prime}^2 + \Gamma_{Z^\prime}^2 \right) }  \\
=  \frac{\alpha^2}{\pi^2}\frac{\left(n_R - n_L \right)^2 (g^\prime)^4 M_\chi^2  s }{ M_{Z^\prime}^2 \left( M_{Z^\prime}^2 + \Gamma_{Z^\prime}^2 \right)  } 
\left( \sum_f  g_A^f N_c^f Q_f^2 \left[1 + 2 M_f^2 C_0 (0,0,s;M_f,M_f,M_f)\right] \right)^2. 
\end{multline}
The cross section times velocity for this channel in the non-relativistic limit is given by
\begin{align}
 \sigma(\bar{\chi} \chi \to \gamma \gamma) v &= \frac{   \left|\sum_f \left(  M_\chi^2 A_7^f - \frac{1}{2} A_1^f  \right) \right|^2 \left(n_L - n_R \right)^2 (g^\prime)^2 M_\chi^2 }
 { 4 \pi \, M_{Z^\prime}^2 \left( M_{Z^\prime}^2 + \Gamma_{Z^\prime}^2 \right)   }.
\end{align}

\item ${\bar{\chi} \chi \to h \gamma}$: 
For the dark matter annihilation into the SM Higgs and a photon one finds the amplitude 
\begin{multline}
\left| \overline{\mathcal{M}} ({\bar{\chi} \chi \rightarrow h \gamma }) \right|^2 = \frac{ (g^\prime)^2 \left| C_{Z^\prime h \gamma} \right|^2 \left(M_h^2-s\right)^2 }{4 \left[ \left( s - M_{Z^\prime}^2 \right)^2 +  M_{Z^\prime}^2  \Gamma_{Z^\prime}^2 \right]} \\
\times \left[ \cos^2\theta \ (n_L^2+ n_R^2) (s - 4 M_\chi^2) + s (n_L^2 + n_R^2) + 8 M_\chi^2 n_L n_R  \right],
\end{multline}
where $\theta$ is the angle between $h$ and $\gamma$ in the center-of-mass system. The corresponding annihilation cross section is given by
\begin{align}
 \sigma(\bar{\chi} \chi \to h \gamma) &= \frac{(g^\prime)^2 | C_{ Z^\prime h \gamma}  |^2 (s- M_h^2)^3 }{48 \pi s^{3/2} } \frac{\left[(n_L^2 + n_R^2) s - (n_L^2 - 6 n_L n_R + n_R^2) 
 M_\chi^2\right]}{ \sqrt{s-4 M_\chi^2}\left[(s- M_{Z^\prime}^2)^2 + \Gamma_{Z^\prime}^2 M_{Z^\prime}^2\right]} .
 \end{align}

 \item ${\bar{\chi} \chi \to Z \gamma}$: 
 In general, the explicit form of the amplitude for the dark matter annihilation into a $Z$ and a photon is very involved and cannot be given here. Here we list the result for $n_L=n_R = n$ where the integrated amplitude is given by
  \begin{multline}
  \int \frac{d \Omega}{2 \pi}\left| \overline{\mathcal{M}} ({\bar{\chi} \chi \rightarrow Z \gamma }) \right|^2 = \frac{(g')^4 g_2^2 e^2 n^2}{768 \pi^4 \cos^2 \theta_W } \\ 
  \times \frac{ \left(2 M_\chi^2+s\right) \left(M_Z^2-s\right)^{2} \left(M_Z^2+s\right)  \left|\sum_f Q_f N_c^f \left( C^f_0 + B_3^f M_Z^2 \right) \right|^2 }
 {M_Z^2 \, s \,\left[(s- M_{Z^\prime}^2)^2 + \Gamma_{Z^\prime}^2 M_{Z^\prime}^2\right] },
 \end{multline}
 while the cross section is given by
 \begin{align}
 & \sigma(\bar{\chi} \chi \to Z \gamma) = \frac{(g')^4 g_2^2 e^2 n^2  \left(2 M_\chi^2+s\right) \left(s-M_Z^2\right)^{3} \left(M_Z^2+s\right)  \left|\sum_f Q_f N_c^f\left( C^f_0 + B_3^f M_Z^2\right) \right|^2 }
 {   24576 \pi^5 \cos\theta_W^2 \,  M_Z^2 \, s^{5/2} \, \sqrt{s-4 M_\chi^2 }    \left[(s- M_{Z^\prime}^2)^2 + \Gamma_{Z^\prime}^2 M_{Z^\prime}^2\right] }.
 \label{chichiZgamma}
 \end{align}
 
\end{itemize}

Notice that these results are very general and can be used for any dark matter model with the features discussed above. For simplicity, we show the results 
for the ${\bar{\chi} \chi \to Z \gamma}$ only for a vector coupling of the $Z'$ to dark matter. 
In Refs.~\cite{Jackson:2009kg,Jackson:2013pjq,Jackson:2013tca}, some of these effective couplings have been computed and the implications for dark matter models have been investigated in detail. 

\subsection{Gamma-Ray Lines and \texorpdfstring{$B-L$}{B-L} Symmetry}
In the simple $B-L$ dark matter model one has only vector couplings to the $B-L$ gauge boson and there are only two relevant channels for the annihilation into gamma-ray lines:
$$\bar{\chi} \chi \  \to \  Z_{BL}^* \  \to \  h \gamma, Z \gamma.$$
The energy of the line signal for the process $\bar{\chi} \chi \to \gamma X$ is given by
\begin{equation}
 E_\gamma = M_\chi \left( 1-\frac{m_X^2}{4 M_\chi^2}\right),
\end{equation}
where $m_X$ is the mass of the particle $X$. The cross section for the dark matter annihilation into the Standard Model Higgs and a photon is given by
\begin{align}
 \sigma(\bar{\chi} \chi \to h \gamma)  = \frac{n^2 g_{BL}^2 | C_{ Z_{BL} h \gamma}  |^2 (s- M_h^2)^3 }{24 \pi s^{3/2} } \frac{\left(s + 2 M_\chi^2\right)}
 { \sqrt{s-4 M_\chi^2} \left[(s - M_{Z_{BL}}^2)^2 + \Gamma_{Z_{BL}}^2 M_{Z_{BL}}^2\right]} .
\end{align} 
The explicit expression for the one-loop generated coupling $C_{ Z_{BL} h \gamma}$ can be found using Eq.~\eqref{eq:hgammaCoefficient} where $g_V^t$ is replaced by the $B-L$ number 1/3 for the top quark and $g^\prime \to g_{BL}$. 
In the non-relativistic limit the above cross section times velocity reads as
\begin{align}
 \sigma(\bar{\chi} \chi \to h \gamma) v = \frac{n^2 g_{BL}^2 | C_{ Z_{BL} h \gamma} |^2 ( 4 M_\chi^2 - M_h^2)^3 }
 {32 \pi M_\chi^2 \left[(M_{Z_{BL}}^2 - 4 M_\chi^2 )^2 + \Gamma_{Z_{BL}}^2 M_{Z_{BL}}^2\right]} .
\end{align}
The annihilation cross section for $\bar{\chi} \chi \  \to \  Z_{BL}^* \  \to \  Z \gamma$ is given by Eq.~\eqref{chichiZgamma}.

\begin{figure}[t]
 \includegraphics[width=0.48\linewidth]{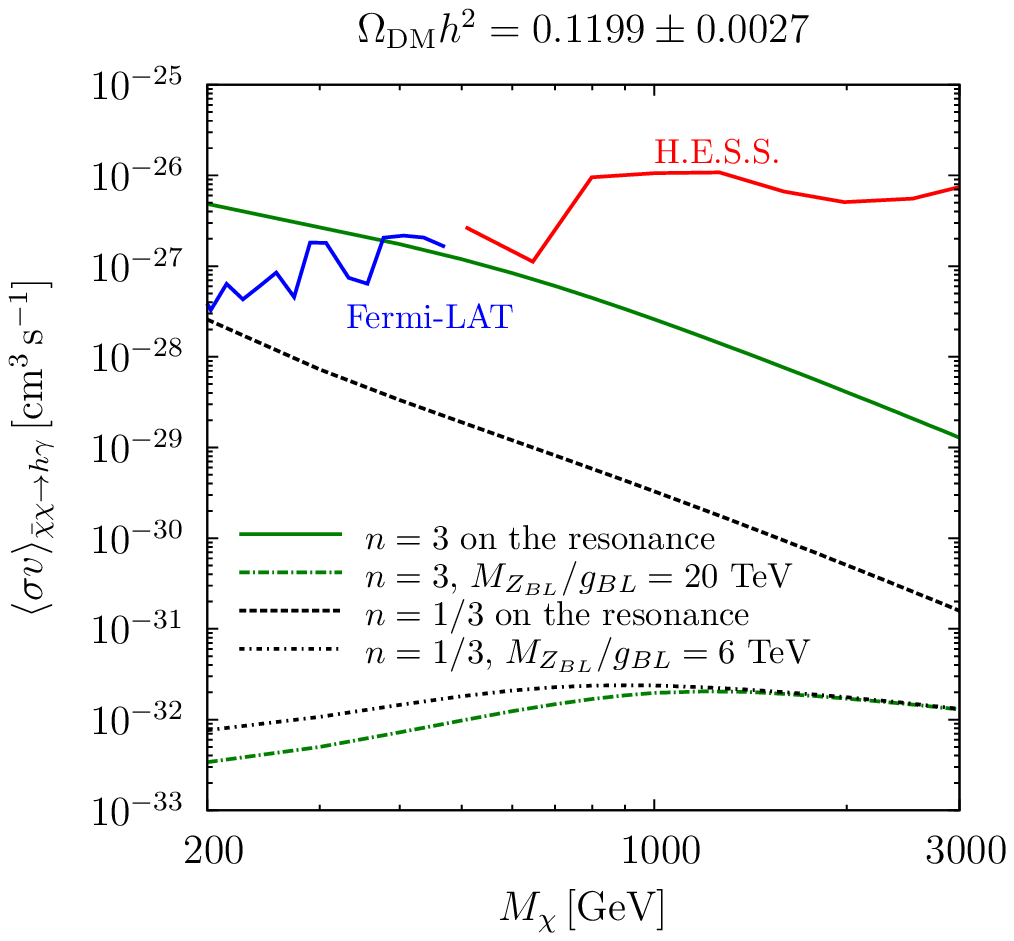}
 \includegraphics[width=0.48\linewidth]{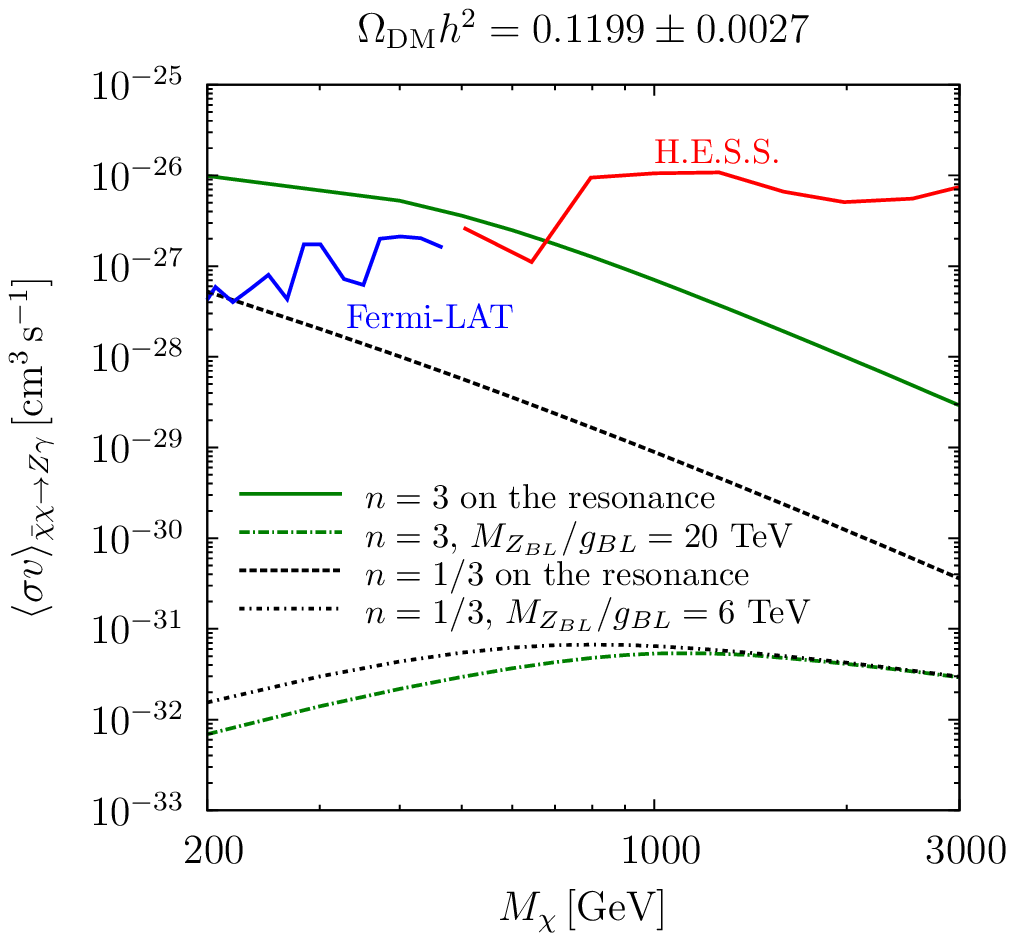}
 \caption{Allowed parameter space for the indirect detection for the channels $\bar{\chi} \chi \to h \gamma$ (left panel) and $\bar{\chi} \chi \to Z \gamma$ (right panel) compatible with the relic density constraint. The limits of Fermi-LAT~\cite{TheFermi-LAT:2015gja} and H.E.S.S.~\cite{Abramowski:2013ax} are shown by the blue and red lines, respectively. See Tab.~\ref{tab:limits} in Appendix~\ref{app:limits} for the values of the cross section limits. Note that the highest cross section in the indirect detection experiments corresponds to the parameters leading to the lowest cross sections in the direct searches. \label{fig:indirectDetection1} } 
\end{figure}

In Fig.~\ref{fig:indirectDetection1} we show the allowed parameter space for the indirect detection for the channels $\bar{\chi} \chi \to h \gamma$ (left panel) and $\bar{\chi} \chi \to Z \gamma$ (right panel) compatible with the relic density constraint. We show the results in two different scenarios for the dark matter charge $n=3$ and $n=1/3$. Only in the case $n=3$ the bounds from Fermi-LAT rule out a small part of the parameter space in the low mass region. In both panels we show the full range for the annihilation cross sections $\langle \sigma v \rangle_{\bar{\chi} \chi \to Z \gamma}$ and $\langle \sigma v \rangle_{\bar{\chi} \chi \to h \gamma}$ which is defined by the collider limits and the resonance regions. The complete region between the two curves is allowed in the model.  However, the region close to the resonance can be ruled out in the near future. 

\begin{figure}[t]
 \includegraphics[width=0.48\linewidth]{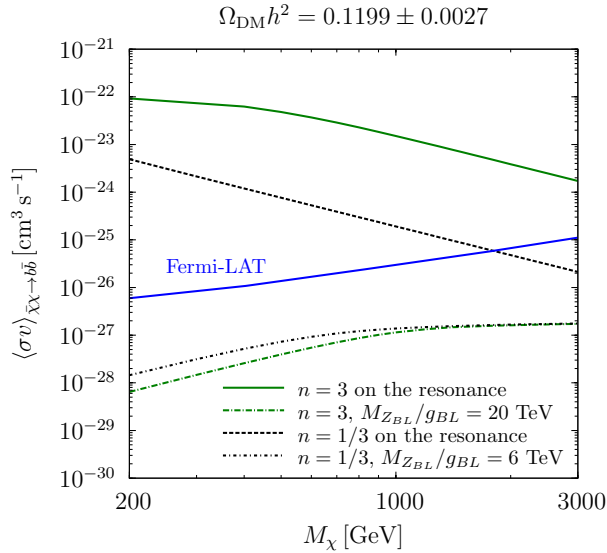}
 \caption{Allowed parameter space for the dark matter annihilation into two bottom quarks compatible with the relic density constraint. The experimental bounds from Fermi-LAT~\cite{Ackermann:2015zua} are given.\label{fig:indirectDetection2} } 
\end{figure}

In Fig.~\ref{fig:indirectDetection2} we show the predictions for the cross section of the dark matter annihilation $\bar{\chi} \chi \to b \bar{b}$. 
Again, we find the full range for the annihilation cross section in agreement with the relic density constraints.
The bounds from Fermi-LAT rule out a significant fraction of the parameter space in both scenarios. As one expects, 
this bound is much stronger than the bounds from gamma-ray lines. It is important to mention that in the resonance 
region the direct detection limits are irrelevant and the best way to probe the model is through indirect detection.

We have investigated carefully the final state radiation $\bar{\chi} \chi \to \bar{f} f \gamma$, which is mediated by the $Z_{BL}$ at tree level and contributes to the continuum gamma-ray spectrum. 
As we have discussed above, the gamma-ray line in this model is possible only at the quantum level and the cross sections are much smaller than the final state radiation. 
Therefore, it is very challenging to distinguish the gamma-ray lines from the continuum spectrum. 
For example, when $n=1/3$ and $M_\chi = \unit[200]{GeV}$, the difference between the gamma-ray lines and the continuum is only about one percent.
Let us stress that the possibility to distinguish the gamma-ray lines from the continuum is a requirement to actually use experimental limits from line searches to derive bounds on a particular dark matter model. 
In the future one could have a very good energy resolution in experiments such as Gamma-400~\cite{G400-1,G400-2,G400-3} and one can investigate this issue in more detail.

\section{Summary}
\label{sec:summary}
We have studied the relic density constraints and the predictions for the direct detection experiments in simplified models for Dirac dark matter. 
In these models the dark matter properties are defined mainly by the gauge interactions, and one can understand dynamically the dark matter stability.
We discussed the cases where one uses the Stueckelberg or the Higgs mechanism for generating the gauge boson mass. 
We have presented general results for the three possible annihilation channels for the Dirac dark matter giving rise to gamma-ray lines, $\bar{\chi} \chi \to \gamma \gamma, h \gamma, Z \gamma$. 
We have shown that the channel $\bar{\chi} \chi \to \gamma \gamma$ is present only when the $Z^\prime$ has an axial coupling 
to fermions inside the loop generating the effective coupling $Z^\prime \gamma \gamma$. The channel $\bar{\chi} \chi \to h \gamma$ is mainly generated by the top quark because
today one cannot have heavy chiral fermions which would change the Higgs properties. These results can be used for any model with a Dirac dark matter charged 
under a new Abelian force.

In order to illustrate our results numerically we investigated a simple model based on local $B-L$ where the neutrino masses could be generated through the seesaw 
mechanism and the dark matter is a Dirac fermion charged under $B-L$. In this case there are only two annihilation channels, $\bar{\chi} \chi \to  h \gamma, Z \gamma$.
We have shown the numerical predictions for these channels taking into account the relic density and collider constraints. We have investigated the correlation 
between the Fermi-LAT and H.E.S.S.\ bounds on gamma-ray lines and the annihilation into bottom quarks to show the constraints on these models. The results presented in this paper tell us how much the bounds on gamma-ray lines 
must be improved  to be able to rule out or test some well-motivated and simple dark matter models.

\section*{Acknowledgments}
P.F.P.\ thanks S.\ Profumo for discussions. We thank Hiren H.\ Patel for helpful discussions and support while using Package-X~\cite{Hiren} for the calculation of one-loop integrals and for pointing out an issue in the calculation of the $\gamma \gamma$ channel.

\appendix

\section{Experimental Constraints on the Annihilation Cross Sections}
\label{app:limits}
Limits on the cross sections are set using the experimental limit on the photon flux, and the relations are different for the various annihilation channels considered in this article. 
\begin{itemize}
\item $\bar{\chi} \chi \rightarrow \gamma \gamma$: In the case of the annihilation of a Dirac dark matter into two photons the cross section is related to the flux by
\begin{equation}
\langle \sigma v \rangle_{\bar{\chi}\chi \to \gamma \gamma} =  \frac{8 \pi}{J_\text{ann}} E_\gamma^2 \,\Phi_\gamma\,.
\end{equation}
\item $\bar{\chi} \chi \rightarrow  X \gamma$: When one has the annihilation into a particle with mass $m_X$ and a photon, the relation is 
\begin{equation}
\langle \sigma v \rangle_{\bar{\chi}\chi \to X \gamma} = \frac{4 \pi}{ J_\text{ann} }  \left(E_\gamma +\sqrt{E_\gamma^2 + m_X^2}\right)^2 \, \Phi_\gamma \,, 
\end{equation}
where $E_\gamma = M_\chi \left( 1- \frac{m_X^2}{4 M_\chi^2} \right)$ has been used.
\end{itemize}
With the help of these relations, one can translate the limits from Fermi-LAT~\cite{TheFermi-LAT:2015gja} and H.E.S.S.~\cite{Abramowski:2013ax} given for the annihilation into two gammas into the $h \gamma$ and $Z \gamma$ channels. See Tab.~\ref{tab:limits} for the Fermi-LAT limits in the range $ \unit[30]{GeV} < E_\gamma < \unit[500]{GeV}$.

\begin{table}
\caption{Limits on the gamma-ray lines from the Fermi-LAT collaboration for $E_\gamma > \unit[30]{GeV}$ for the R3 region with a NFWc dark matter profile~\cite{TheFermi-LAT:2015gja}. Using the limits on the $\gamma \gamma$ channel, we derive the corresponding cross section limits and dark matter masses for the $Z \gamma$ and $h \gamma$ channels. Notice that for the Dirac dark matter case the cross section limits have to be multiplied by a factor of two. \label{tab:limits}}
\begin{tabular}{cc|cc|cc}
\hline\hline
\multicolumn{2}{c}{$\gamma\gamma$ channel} & \multicolumn{2}{c}{$h\gamma$ channel} &  \multicolumn{2}{c}{$Z\gamma$ channel} \\
$E_\gamma$ [GeV] & $\langle \sigma v \rangle_{\gamma \gamma}$ [$\unit[10^{-29}]{cm^3/s}$] & $M_\chi$ [GeV] & $\langle \sigma v \rangle_{h \gamma}$ [$\unit[10^{-29}]{cm^3/s}$] & $M_\chi$ [GeV] & $\langle \sigma v \rangle_{Z \gamma}$ [$\unit[10^{-29}]{cm^3/s}$] \\ \hline
31.2 & 0.661 & 80.4 & 8.77 & 63.8 & 5.53 \\
33.0 & 0.695 & 81.5 & 8.47 & 65.0 & 5.39 \\
34.9 & 1.42 & 82.7 & 15.9 & 66.3 & 10.2 \\
36.9 & 3.08 & 84.0 & 31.9 & 67.6 & 20.7 \\
39.0 & 3.87 & 85.3 & 37.0 & 69.1 & 24.3 \\
41.3 & 3.79 & 86.8 & 33.5 & 70.7 & 22.2 \\
43.8 & 6.41 & 88.5 & 52.3 & 72.5 & 35.1 \\
46.4 & 6.54 & 90.2 & 49.4 & 74.4 & 33.6 \\
49.1 & 5.62 & 92.0 & 39.5 & 76.3 & 27.2 \\
52.1 & 3.12 & 94.1 & 20.3 & 78.6 & 14.2 \\
55.2 & 3.38 & 96.2 & 20.5 & 80.9 & 14.5 \\
58.6 & 7.13 & 98.6 & 40.4 & 83.5 & 29.0 \\
62.2 & 6.95 & 101 & 36.8 & 86.3 & 26.8 \\
66.0 & 4.59 & 104 & 22.8 & 89.3 & 16.8 \\
70.1 & 5.18 & 107 & 24.1 & 92.6 & 18.1 \\
74.5 & 5.56 & 110 & 24.4 & 96.1 & 18.5 \\
79.2 & 3.08 & 114 & 12.7 & 100 & 9.82 \\
84.2 & 2.87 & 118 & 11.2 & 104 & 8.78 \\
89.6 & 2.87 & 122 & 10.6 & 109 & 8.45 \\
95.4 & 2.82 & 127 & 9.93 & 114 & 8.01 \\
102 & 5.77 & 132 & 19.3 & 119 & 15.8 \\
108 & 5.73 & 137 & 18.4 & 125 & 15.3 \\
115 & 15.2 & 143 & 46.8 & 131 & 39.4 \\
123 & 15.1 & 149 & 44.6 & 138 & 38.0 \\
131 & 10.8 & 156 & 30.7 & 145 & 26.6 \\
140 & 5.29 & 164 & 14.5 & 154 & 12.7 \\
150 & 10.6 & 173 & 28.2 & 163 & 25.0 \\
160 & 8.15 & 182 & 21.0 & 172 & 18.9 \\
171 & 13.0 & 192 & 32.6 & 182 & 29.6 \\
183 & 6.68 & 203 & 16.4 & 194 & 15.0 \\
196 & 13.3 & 214 & 31.8 & 206 & 29.4 \\
210 & 9.19 & 227 & 21.5 & 219 & 20.1 \\
225 & 13.0 & 241 & 29.9 & 234 & 28.1 \\
241 & 18.7 & 256 & 42.3 & 249 & 40.0 \\
259 & 10.2 & 273 & 22.7 & 267 & 21.6 \\
276 & 41.2 & 290 & 90.7 & 283 & 86.8 \\
294 & 41.4 & 307 & 90.2 & 301 & 86.7 \\
321 & 17.3 & 333 & 37.2 & 327 & 36.0 \\
345 & 15.0 & 356 & 32.0 & 351 & 31.0 \\
367 & 48.7 & 377 & 103 & 373 & 100 \\
396 & 51.7 & 406 & 109 & 401 & 106 \\
427 & 49.6 & 436 & 103 & 432 & 101 \\
462 & 39.4 & 470 & 81.7 & 466 & 80.3 \\
\hline\hline
\end{tabular}
\end{table}

\section{Loop Functions}\label{app:loopFunctions}

In this appendix, we define the loop functions used throughout the article.
The function $\Lambda(s,m_1,m_2)$ contains the logarithmic discontinuity of the Passarino--Veltman $B_0$ function. For the special case of $m_1=m_2=m$ it is given by
\begin{equation}
\Lambda(s, m, m) = \sqrt{1-\frac{4 m^2}{s}} \ln \left[\frac{2 m^2}{2 m^2-s \left(\sqrt{1-\frac{4 m^2}{s}}+1\right)}\right].
\end{equation}
For special cases, the Passarino--Veltman $C_0$ function can be given as
\begin{equation}
 C_0(0, 0, s; m, m, m) = \frac{1}{2s} \ln^2 \left( \frac{\sqrt{1-\frac{4 m^2}{s}}-1}{\sqrt{1-\frac{4 m^2}{s}}+1} \right)
\end{equation}
and
\begin{align}
 C_0(0,M^2,s;m,m,m) = \frac{1}{2 (M^2 - s)} \left[ \ln^2 \left( \frac{\sqrt{1-\frac{4 m^2}{M^2}}-1}{\sqrt{1-\frac{4 m^2}{M^2}}+1} \right) - \ln^2 \left( \frac{\sqrt{1-\frac{4 m^2}{s}}-1}{\sqrt{1-\frac{4 m^2}{s}}+1} \right)\right].
\end{align}



\end{document}